\documentclass[aps,prb,reprint,superscriptaddress,floatfix]{revtex4-2}

\usepackage{graphicx}
\usepackage{amsmath,amssymb,amsfonts}
\usepackage{bm}

\begin{document}

\title{Three-dimensional Ising superconductors designed via inversion-symmetry breaking in intercalated NbSe$_2$ and NbTe$_2$}

\author{Wenqian Tu}
\affiliation{Key Laboratory of Materials Physics, Institute of Solid State Physics, HFIPS, Chinese Academy of Sciences, Hefei 230031, China}
\affiliation{University of Science and Technology of China, Hefei 230026, China}

\author{Run Lv}
\affiliation{Key Laboratory of Materials Physics, Institute of Solid State Physics, HFIPS, Chinese Academy of Sciences, Hefei 230031, China}
\affiliation{University of Science and Technology of China, Hefei 230026, China}

\author{Xiaoying Li}
\affiliation{Key Laboratory of Materials Physics, Institute of Solid State Physics, HFIPS, Chinese Academy of Sciences, Hefei 230031, China}
\affiliation{University of Science and Technology of China, Hefei 230026, China}

\author{Li'e Liu}
\affiliation{Key Laboratory of Materials Physics, Institute of Solid State Physics, HFIPS, Chinese Academy of Sciences, Hefei 230031, China}
\affiliation{University of Science and Technology of China, Hefei 230026, China}

\author{Dingfu Shao}
\affiliation{Key Laboratory of Materials Physics, Institute of Solid State Physics, HFIPS, Chinese Academy of Sciences, Hefei 230031, China}

\author{Yuping Sun}
\affiliation{High Magnetic Field Laboratory, HFIPS, Chinese Academy of Sciences, Hefei 230031, China}
\affiliation{Key Laboratory of Materials Physics, Institute of Solid State Physics, HFIPS, Chinese Academy of Sciences, Hefei 230031, China}
\affiliation{Collaborative Innovation Center of Microstructures, Nanjing University, Nanjing 210093, China}

\author{Wenjian Lu}
\email{wjlu@issp.ac.cn}
\affiliation{Key Laboratory of Materials Physics, Institute of Solid State Physics, HFIPS, Chinese Academy of Sciences, Hefei 230031, China}

\date{\today}

\begin{abstract}
Ising superconductors exhibit in-plane upper critical fields far exceeding the Pauli paramagnetic limit, a hallmark first established in two-dimensional (2D) monolayer transition-metal dichalcogenides (TMDs). This field resilience requires the coexistence of strong spin-orbit coupling (SOC) and broken inversion symmetry, yet three-dimensional (3D) bulk realizations remain scarce because equilibrium stacking typically restores inversion symmetry. Here we demonstrate that intercalation provides a practical route to break this symmetry, systematically designing 16 NbSe$_2$- and NbTe$_2$-based compounds from four intercalants (In, Sn, Pb, Bi) across two polytypes: non-centrosymmetric $P\bar{6}m2$ and centrosymmetric $P6_3/mmc$. Four compounds in the $P\bar{6}m2$ phase, InNbSe$_2$, SnNbSe$_2$, PbNbSe$_2$, and PbNbTe$_2$, emerge as promising 3D Ising superconductors. They exhibit SOC splittings of 80-100 meV near the Fermi level, dominant out-of-plane spin polarization, and anisotropic superconductivity with $T_c=2.6$-$5.4$ K. Notably, spin-texture analysis reveals that the efficiency of Ising protection is governed not by the magnitude of SOC splitting alone but by the out-of-plane spin purity on the Fermi surface. Bogoliubov-de Gennes (BdG) calculations predict in-plane upper critical fields reaching 4-7 times the Pauli limit. These findings establish intercalation as a promising symmetry-engineering strategy for realizing 3D Ising superconductors in TMDs.
\end{abstract}

\maketitle

\section{Introduction}
\label{sec:intro}

Ising superconductivity originates from spin-valley locking, in which electron spins are pinned out of plane with opposite orientations at time-reversed valleys. In transition-metal dichalcogenides (TMDs), this locking emerges when broken inversion symmetry acts together with strong spin-orbit coupling (SOC) to generate an effective Zeeman field along the out-of-plane direction \cite{saito2016,wickramaratne2023,wang2021,qiu2021}. Because Cooper pairs formed from these time-reversed valley states carry antiparallel out-of-plane spins, they remain protected against in-plane magnetic fields, allowing the in-plane upper critical field $H_{c2,\parallel}$ to far exceed the Pauli paramagnetic limit $H_P$ \cite{clogston1962}. This hallmark was first established in atomically thin two-dimensional (2D) TMDs such as 1$H$-NbSe$_2$ \cite{ugeda2015}, 1$H$-TaS$_2$ \cite{lian2022,delabarrera2018}, and gated MoS$_2$ \cite{lu2015}. In these systems, the combination of broken inversion symmetry and strong SOC originating from the heavy transition-metal $d$ orbitals gives rise to robust spin-valley locking.

Yet 2D Ising superconductors inherit the fragility of atomic-scale materials, whereas three-dimensional (3D) bulk crystals offer greater robustness, scalability, and practical utility. In most bulk TMDs, the equilibrium layer stacking typically restores inversion symmetry and thereby extinguishes the antisymmetric SOC required for spin-valley locking, rendering 3D Ising superconductors exceedingly rare. The canonical example is 2$H$-NbSe$_2$, whose AB stacking eliminates the Ising protection present in the monolayer \cite{xi2015}. A notable exception is bulk 4$H_a$-NbSe$_2$, a naturally non-centrosymmetric polytype ($P\bar{6}m2$), which exhibits $H_{c2,\parallel}$ nearly three times $H_P$ \cite{patra2025,volavka2026}, demonstrating that 3D Ising superconductivity becomes accessible once the stacking symmetry is broken.

A more general and tunable strategy is intercalation: inserting guest atoms into the van der Waals (vdW) gap breaks inversion symmetry while preserving the 3D layered structure. This approach has already borne fruit experimentally: 3D Ising superconductivity was reported in as-grown bilayer-Sn-intercalated TaSe$_2$ \cite{zheng2025} and in Li-intercalated NbSe$_2$ \cite{ji2024}, with $H_{c2,\parallel}$ exceeding $H_P$ by factors of roughly 2.6 and 2, respectively. On the host side, 2$H$-NbTe$_2$, the heavier chalcogen analogue of NbSe$_2$, was recently synthesized \cite{jin2024}, extending the intercalation platform to tellurides. Despite these advances, only a handful of intercalant-host combinations have been explored, and a systematic screening across intercalant chemistry and stacking polytypes, together with a quantitative account of what actually controls the Ising protection efficiency, remains conspicuously absent.

In this work, we systematically design 16 intercalated compounds spanning a broad chemical space: four intercalants (In, Sn, Pb, Bi) chosen to cover wide ranges of atomic mass and ionic radius, inserted into NbSe$_2$ and NbTe$_2$ hosts. Each composition is constructed in both the non-centrosymmetric $P\bar{6}m2$ polytype (adopted by PbTaSe$_2$ \cite{bian2016}) and its centrosymmetric $P6_3/mmc$ counterpart, enabling direct symmetry control at fixed composition. After screening for dynamical stability, electronic structure, and superconductivity, we identify four viable 3D Ising superconductor candidates, InNbSe$_2$, SnNbSe$_2$, PbNbSe$_2$, and PbNbTe$_2$, all in the non-centrosymmetric $P\bar{6}m2$ phase. These compounds exhibit SOC band splittings of 80-100 meV and superconducting transition temperatures $T_c$ of 2.6-5.4 K. The splitting derives mostly from Nb-$4d$ states and varies only weakly across the series; remarkably, however, the resulting $H_{c2,\parallel}$, predicted by Bogoliubov-de Gennes (BdG) calculations to reach 4-7 times $H_P$, does not simply track the SOC splitting strength. Instead, out-of-plane spin purity on the Fermi surface (FS) governs the Ising protection efficiency: residual in-plane spin components provide a parasitic depairing channel under in-plane fields. Anisotropic Eliashberg calculations further reveal highly anisotropic superconductivity in all four candidates, with pronounced two-gap character in SnNbSe$_2$ and PbNbSe$_2$. Together, these results establish intercalation as a promising design strategy for realizing 3D Ising superconductors in TMDs.

\section{Methods}
\label{sec:methods}

First-principles calculations based on density functional theory (DFT) were performed using the Vienna \textit{ab initio} Simulation Package (VASP) \cite{kresse1996,kresse1999}. The exchange-correlation functional was treated within the generalized gradient approximation (GGA) of Perdew-Burke-Ernzerhof (PBE) \cite{perdew1996}. SOC was included in all calculations. A plane-wave kinetic energy cutoff of 500 eV was adopted, and structural relaxations were carried out until the residual forces fell below $10^{-4}$ eV/\AA, with vdW interactions treated via the DFT-D3 scheme \cite{grimme2011}. The Brillouin zone (BZ) was sampled by a $\Gamma$-centered grid with a resolution of $0.02$ \AA$^{-1}$ for self-consistent calculations. Spin textures on the FS were visualized using the IFermi package \cite{ganose2021}.

Maximally localized Wannier functions (MLWF) were constructed with Wannier90 \cite{mostofi2014} using Nb-$4d$, chalcogen-$p$, and intercalant-$p$ orbitals as initial projections. The resulting real-space tight-binding Hamiltonian $H(\mathbf{R})$ was subsequently used to calculate the BdG superconducting quasiparticle density of states (DOS). 3D band structures and topological surface states were calculated using WannierTools \cite{wu2018} from the Wannier90-derived Hamiltonian.

Phonon dispersions were calculated using the QUANTUM ESPRESSO package \cite{giannozzi2009} with density functional perturbation theory (DFPT) \cite{baroni2001}. A $6\times6\times2$ q-point grid and a $12\times12\times4$ k-point grid were employed to sample the BZ. The wave function and charge density energy cutoffs were set to 100 Ry and 600 Ry, respectively. The Gaussian smearing method with a smearing parameter of $\sigma=0.02$ Ry was used. Systems exhibiting imaginary phonon frequencies at any q-point were classified as dynamically unstable and excluded from further analysis. The superconducting properties of the remaining 11 stable systems were calculated using the EPW code \cite{ponce2016,lee2023}, which interpolates the DFPT electron-phonon coupling (EPC) matrix elements onto $36\times36\times12$ q- and $72\times72\times24$ k-point grids via MLWF. The $T_c$ was calculated from the Allen-Dynes-modified McMillan formula \cite{allen1975,dynes1972}:
\begin{equation}
T_c = \frac{\omega_{\log}}{1.2}\exp\left[-\frac{1.04(1+\lambda)}{\lambda-\mu^{*}(1+0.62\lambda)}\right],
\end{equation}
where the total EPC constant is
\begin{equation}
\lambda = 2\int \frac{\alpha^2 F(\omega)}{\omega}\,d\omega .
\end{equation}
The Coulomb pseudopotential $\mu^{*}$ is generally assumed to be the typical value of 0.1 \cite{calandra2011,kortus2001}. Here, $\omega_{\mathbf{q}\nu}$ is the phonon frequency, $\lambda_{\mathbf{q}\nu}$ is the EPC constant contributed by the $\nu$th mode at the wave vector $\mathbf{q}$, defined as
\begin{equation}
\lambda_{\mathbf{q}\nu} = \frac{\gamma_{\mathbf{q}\nu}}{\pi\hbar N(E_F)\,\omega_{\mathbf{q}\nu}^{2}} .
\end{equation}
The logarithmic average frequency $\omega_{\log}$ is defined as
\begin{equation}
\omega_{\log} = \exp\left[\frac{2}{\lambda}\int d\omega\, \frac{\alpha^2 F(\omega)}{\omega}\,\log\omega\right],
\end{equation}
with the Eliashberg spectral function
\begin{equation}
\alpha^2 F(\omega) = \frac{1}{2\pi N(E_F)}\sum_{\mathbf{q}\nu} \frac{\gamma_{\mathbf{q}\nu}}{\hbar\omega_{\mathbf{q}\nu}}\,\delta(\omega-\omega_{\mathbf{q}\nu}) .
\end{equation}
The anisotropic Migdal-Eliashberg (ME) equations were additionally solved within the Fermi-surface-restricted (FSR) approximation \cite{margine2013} to obtain the momentum-resolved superconducting gap.

\section{Results and discussion}
\label{sec:results}

\begin{figure*}[t]
\centering
\includegraphics[width=0.85\linewidth]{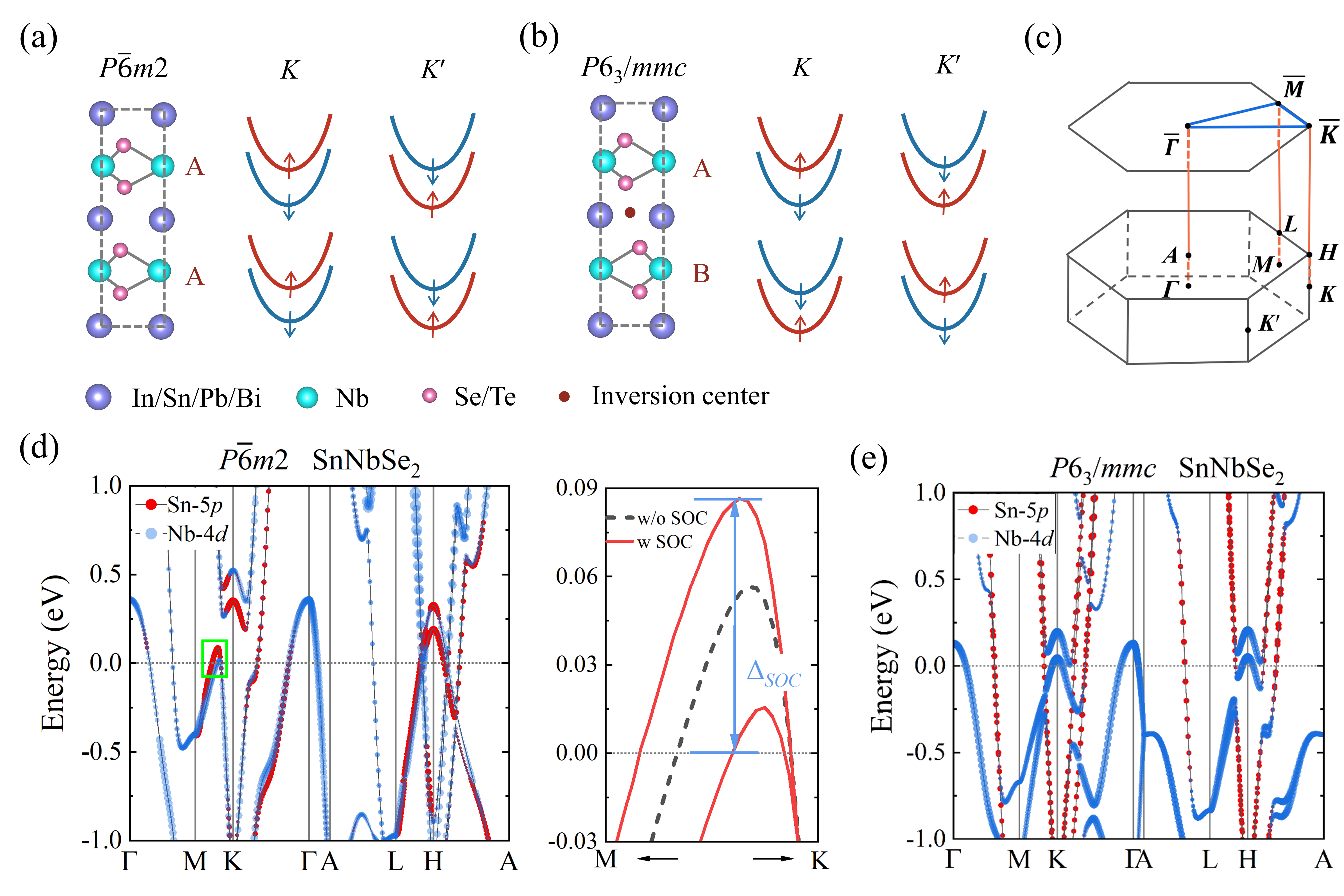}
\caption{(a), (b) Crystal structures and schematic single-layer band dispersions near the $K$ and $K'$ valleys for the non-centrosymmetric $P\bar{6}m2$ phase and the centrosymmetric $P6_3/mmc$ phase. The red dot in (b) marks the inversion center. (c) Bulk and projected (001) surface Brillouin zones of the hexagonal lattice. (d) Orbital-projected band structure of SnNbSe$_2$ in the $P\bar{6}m2$ phase, where the contributions of Sn-$5p$ and Nb-$4d$ orbitals are indicated by red and blue dots, respectively. The right panel enlarges the green-boxed region along $M$-$K$ with (solid red) and without (dashed black) SOC. The blue arrow marks the SOC-induced band splitting $\Delta_{\mathrm{SOC}}$. (e) Orbital-projected band structure of SnNbSe$_2$ in the $P6_3/mmc$ phase.}
\label{fig:structures}
\end{figure*}

We begin by establishing the structural landscape and systematically narrowing the candidate pool through stability and superconductivity screening. The two polytypes investigated here are illustrated in Figs.~\ref{fig:structures}(a) and (b). Both derive from the prototypical 2$H$ stacking of NbSe$_2$/NbTe$_2$ layers, in which each Nb atom is trigonal-prismatically coordinated by six chalcogen atoms, forming Se-Nb-Se (Te-Nb-Te) sandwich layers. Intercalant atoms (In, Sn, Pb, Bi) occupy the octahedral sites of the vdW gap between adjacent NbSe$_2$/NbTe$_2$ layers, each sitting directly above (or below) an Nb atom along the $c$ axis. Fig.~\ref{fig:structures}(c) shows the bulk BZ of the hexagonal lattice with the high-symmetry points labeled, together with the projected (001) surface BZ relevant to the surface-state calculations.

The two polytypes differ fundamentally in their stacking sequence and resulting space-group symmetry. In the $P\bar{6}m2$ polytype (point group $D_{3h}$), the NbSe$_2$/NbTe$_2$ layers stack with identical orientation (AA stacking), and the intercalant occupies a site that preserves the trigonal-prismatic environment while breaking the inversion center along the $c$ axis. This broken inversion symmetry is the necessary condition for Ising-type SOC: the absence of the inversion operation $I:z\to-z$, combined with the preserved $\sigma_h$ mirror symmetry, generates an effective out-of-plane Zeeman field $B_{\mathrm{SOC}}(\mathbf{k})\propto\hat{\mathbf{z}}\times\hat{\mathbf{k}}$ at the $K$ and $K'$ valleys of the hexagonal BZ, with opposite signs at time-reversed valleys \cite{xiao2012}. As illustrated in the band schematic of Fig.~\ref{fig:structures}(a), each NbSe$_2$/NbTe$_2$ monolayer hosts this Ising SOC splitting intrinsically, and because all layers share the same orientation, their spin splittings add constructively, and the out-of-plane spin polarization of the individual layers survives in the bulk.

In contrast, the $P6_3/mmc$ polytype (point group $D_{6h}$) features AB stacking with a restored inversion center marked by a red dot in Fig.~\ref{fig:structures}(b). Inversion symmetry enforces spin degeneracy at every k-point and forbids the antisymmetric SOC term responsible for spin-valley locking. Crucially, as shown in Fig.~\ref{fig:structures}(b), each layer still hosts the same Ising SOC splitting as in the non-centrosymmetric $P\bar{6}m2$ phase, but adjacent layers are rotated by 180 degrees, which inverts the sign of $B_{\mathrm{SOC}}$ in successive layers at the same valley. The alternating spin polarization cancels, yielding spin-degenerate bulk bands despite the finite SOC splitting within each layer. With identical chemical composition, the key distinction between the two polytypes is thus the presence or absence of an inversion center, enabling a direct comparison of Ising-active and Ising-inactive phases at fixed composition.

\begin{table*}[t]
\caption{Space group (SG), calculated in-plane ($a$, $b$) and out-of-plane ($c$) lattice parameters, dynamical stability, and superconducting parameters for 16 candidate intercalated systems. $\checkmark$ denotes dynamically stable, while $\times$ denotes dynamically unstable (imaginary phonon frequencies).}
\label{tab:table1}
\begin{ruledtabular}
\begin{tabular}{lccccccc}
System & SG & $a,b$ (\AA) & $c$ (\AA) & Stable & $\omega_{\log}$ (K) & $\lambda$ & $T_c$ (K)\\
\hline
InNbSe$_2$ & $P\bar{6}m2$ & 3.41 & 9.17 & \checkmark & 83.5 & 0.78 & 3.7\\
SnNbSe$_2$ & $P\bar{6}m2$ & 3.42 & 9.17 & \checkmark & 78.1 & 0.98 & 5.4\\
PbNbSe$_2$ & $P\bar{6}m2$ & 3.45 & 9.31 & \checkmark & 101.5 & 0.58 & 2.6\\
BiNbSe$_2$ & $P\bar{6}m2$ & 3.47 & 9.16 & \checkmark & 83.6 & 0.18 & 0.0\\
InNbTe$_2$ & $P\bar{6}m2$ & 3.59 & 9.98 & $\times$ & --- & --- & ---\\
SnNbTe$_2$ & $P\bar{6}m2$ & 3.60 & 9.93 & $\times$ & --- & --- & ---\\
PbNbTe$_2$ & $P\bar{6}m2$ & 3.65 & 10.13 & \checkmark & 68.7 & 0.90 & 4.1\\
BiNbTe$_2$ & $P\bar{6}m2$ & 3.63 & 9.93 & \checkmark & 49.4 & 0.32 & 0.1\\
InNbSe$_2$ & $P6_3/mmc$ & 3.40 & 18.40 & $\times$ & --- & --- & ---\\
SnNbSe$_2$ & $P6_3/mmc$ & 3.42 & 18.33 & \checkmark & 83.5 & 0.82 & 4.5\\
PbNbSe$_2$ & $P6_3/mmc$ & 3.45 & 18.65 & \checkmark & 95.5 & 0.53 & 1.7\\
BiNbSe$_2$ & $P6_3/mmc$ & 3.47 & 18.47 & \checkmark & 86.1 & 0.18 & 0.0\\
InNbTe$_2$ & $P6_3/mmc$ & 3.59 & 20.03 & $\times$ & --- & --- & ---\\
SnNbTe$_2$ & $P6_3/mmc$ & 3.60 & 19.84 & $\times$ & --- & --- & ---\\
PbNbTe$_2$ & $P6_3/mmc$ & 3.65 & 20.24 & \checkmark & 81.7 & 0.67 & 2.6\\
BiNbTe$_2$ & $P6_3/mmc$ & 3.63 & 19.85 & \checkmark & 58.9 & 0.26 & 0.0\\
\end{tabular}
\end{ruledtabular}
\end{table*}

\begin{table*}[t]
\caption{Space group (SG), SOC splitting $\Delta_{\mathrm{SOC}}$, superconducting transition temperature $T_c$, Pauli paramagnetic limit $H_P$, and calculated upper critical fields for all seven superconducting intercalated systems and two reference NbSe$_2$ polytypes. $H_P = 1.86\times T_c$ \cite{clogston1962}.}
\label{tab:table2}
\begin{ruledtabular}
\begin{tabular}{lccccccc}
System & SG & $\Delta_{\mathrm{SOC}}$ (meV) & $H_P$ (T) & $H_{c2,\parallel}$ (T) & $H_{c2,\perp}$ (T) & $H_{c2,\parallel}/H_P$ & $H_{c2,\parallel}/H_{c2,\perp}$\\
\hline
InNbSe$_2$ & $P\bar{6}m2$ & 80 & 6.8 & 7.2 & 40 & 5.9 & 5.5\\
SnNbSe$_2$ & $P\bar{6}m2$ & 86 & 10.0 & 6.7 & 70 & 7 & 10.5\\
PbNbSe$_2$ & $P\bar{6}m2$ & 91 & 4.8 & 3.5 & 20 & 4.2 & 5.7\\
PbNbTe$_2$ & $P\bar{6}m2$ & 98 & 7.6 & 5.1 & 30 & 3.9 & 5.9\\
SnNbSe$_2$ & $P6_3/mmc$ & 0 & 8.3 & 8 & 10 & 1.2 & 1.2\\
PbNbSe$_2$ & $P6_3/mmc$ & 0 & 3.1 & 4 & 5 & 1.6 & 1.3\\
PbNbTe$_2$ & $P6_3/mmc$ & 0 & 4.7 & 4.5 & 5 & 1.1 & 1.1\\
1$H$-NbSe$_2$ & $P\bar{6}m2$ & 148[8] & 5.6[6,10] & 4[10] & 35[10] & 6.3 & 8.8\\
4$H_a$-NbSe$_2$ & $P\bar{6}m2$ & 150[12] & 11.5[12,37] & 5.5[12] & 31[12] & 2.8 & 5.6\\
\end{tabular}
\end{ruledtabular}
\end{table*}

Of the 16 initial candidate systems ($4$ intercalants $\times\, 2$ hosts $\times\, 2$ polytypes), five exhibit imaginary phonon frequencies indicating dynamical instability: InNbTe$_2$ in both polytypes, SnNbTe$_2$ in both polytypes, and InNbSe$_2$ in the $P6_3/mmc$ phase. These five unstable systems are excluded from further analysis. The complete phonon dispersions for all 16 systems are provided in supplemental Fig.~S1 ($P\bar{6}m2$) and Fig.~S2 ($P6_3/mmc$). The remaining 11 dynamically stable systems (Table~\ref{tab:table1}) span the full $P\bar{6}m2$ and $P6_3/mmc$ contrast for the NbSe$_2$ host with Sn, Pb, and Bi intercalants, as well as the Pb- and Bi-intercalated NbTe$_2$ systems. With the 11 dynamically stable systems identified, we next screen their superconducting properties. Table~\ref{tab:table1} compiles the logarithmic average phonon frequency $\omega_{\log}$, the EPC constant $\lambda$, and superconducting transition temperature $T_c$ for all 11 systems. Seven compounds exhibit finite $T_c$: four in the $P\bar{6}m2$ phase (InNbSe$_2$, SnNbSe$_2$, PbNbSe$_2$, and PbNbTe$_2$, with $T_c = 2.6$-$5.4$ K) and three in the $P6_3/mmc$ phase (SnNbSe$_2$, PbNbSe$_2$, and PbNbTe$_2$, with $T_c = 1.7$-$4.5$ K). The Bi-intercalated systems in both polytypes are non-superconducting and are excluded from further analysis. Among the seven superconducting systems, a clear trend emerges: the $P\bar{6}m2$-phase systems consistently exhibit higher $T_c$ than their counterparts in the $P6_3/mmc$ structure of the same composition (e.g., SnNbSe$_2$: 5.4 K vs 4.5 K; PbNbSe$_2$: 2.6 K vs 1.7 K). The seven superconducting systems thus identified constitute the pool for subsequent analysis.

Fig.~\ref{fig:structures}(d) presents the orbital-projected band structure of SnNbSe$_2$ in the $P\bar{6}m2$ phase. The electronic structure near the Fermi level $E_F$ is dominated by the Nb-$4d$ orbital, with a minor contribution from the Sn-$5p$ orbital. This orbital hierarchy is a universal feature across all intercalated systems studied here and has an important consequence: the low-energy physics is almost governed by Nb-$4d$ electrons, with the intercalant playing an indirect role. Along the $M$-$K$ path, SOC induces a pronounced splitting of the band: $\Delta_{\mathrm{SOC}} = 86$ meV for SnNbSe$_2$ (Fig.~\ref{fig:structures}(d), right panel). The full set of $\Delta_{\mathrm{SOC}}$ splittings across all $P\bar{6}m2$ systems is compared in Table~\ref{tab:table2}. Among the four superconducting candidates, the splitting ranges from 80 meV (InNbSe$_2$) to 98 meV (PbNbTe$_2$). The narrow spread of $\Delta_{\mathrm{SOC}}$, only $\sim$20\% variation across the four superconducting candidates despite spanning three intercalant species and two chalcogens, reflects the dominant Nb-$4d$ character of the k-point valence states: the intercalant and chalcogen orbitals contribute minimally to these bands, so their SOC enters only indirectly through the modified lattice parameters and charge transfer. By contrast, all variants in the $P6_3/mmc$ structure (Fig.~\ref{fig:structures}(e)) exhibit zero SOC splitting, which is a direct consequence of the restored inversion symmetry. The complete SOC band structures for all 16 candidate systems are shown in supplemental Fig.~S3 ($P\bar{6}m2$) and Fig.~S4 ($P6_3/mmc$).

\begin{figure}[t]
\includegraphics[width=\linewidth]{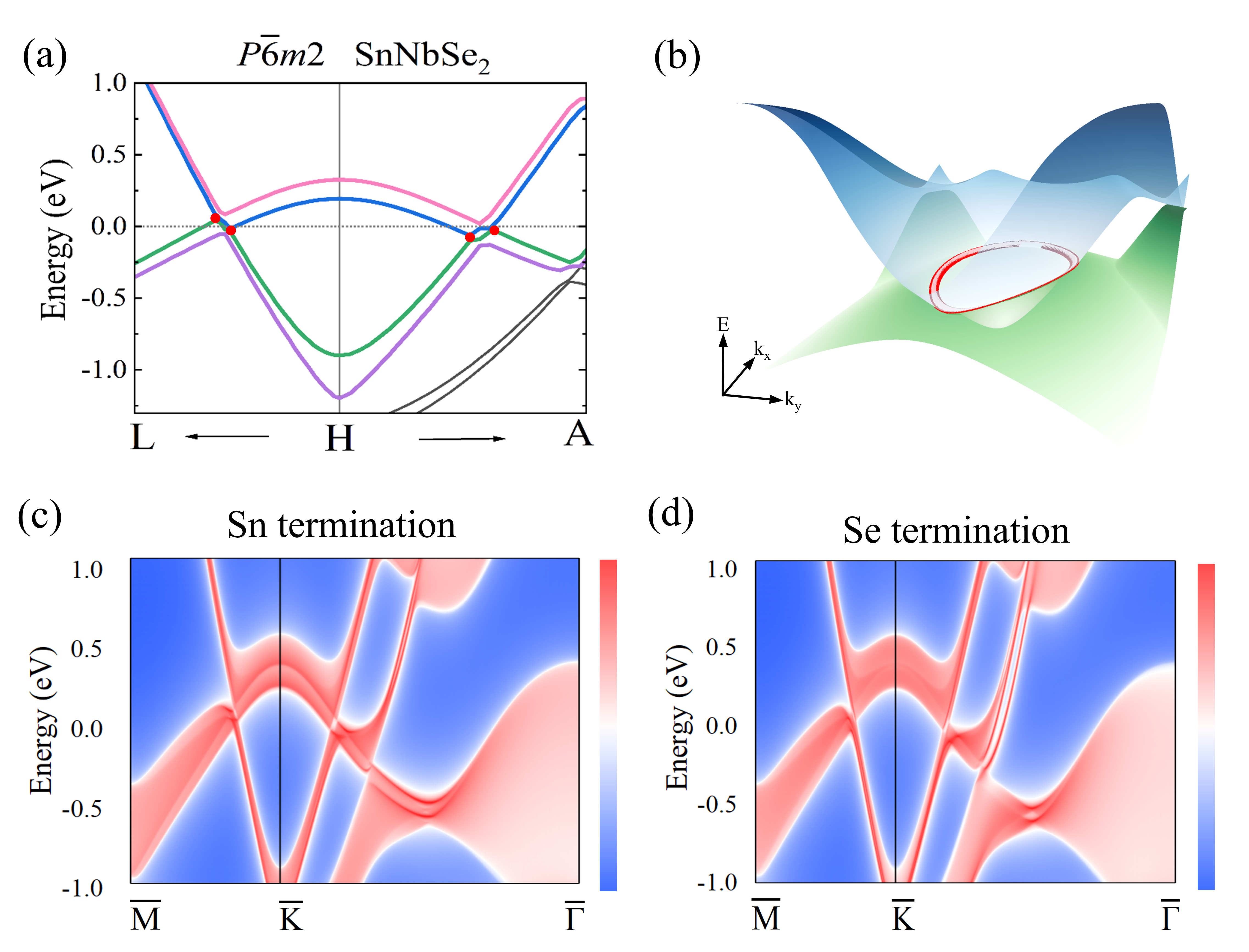}
\caption{(a) Band structure of SnNbSe$_2$ in the $P\bar{6}m2$ phase along the $L$-$H$-$A$ path, with nodal-line crossings marked by red dots. (b) 3D band dispersion in the $k_x$-$k_y$ plane, where nodal rings are highlighted as red loops. (c), (d) Surface state spectra for the Sn-terminated and Se-terminated (001) surfaces, respectively.}
\label{fig:nodal}
\end{figure}

This on-off dichotomy is symmetry-enforced: every $P\bar{6}m2$ compound exhibits finite $\Delta_{\mathrm{SOC}}$ while every $P6_3/mmc$ counterpart shows zero splitting, regardless of intercalant or chalcogen identity (Table~\ref{tab:table2}). The physical origin is transparent: the antisymmetric SOC term $H_{\mathrm{ASOC}}\propto(\hat{\mathbf{z}}\times\hat{\mathbf{k}})\cdot\bm{\sigma}$, which generates the effective out-of-plane Zeeman field responsible for spin-valley locking, is odd under spatial inversion and therefore vanishes identically in centrosymmetric crystals. Consequently, the bands in the centrosymmetric phase remain spin-degenerate at all k-points despite each constituent monolayer possessing its own Ising SOC (Fig.~\ref{fig:structures}(b)). The stacking polytype thus dictates whether Ising SOC is symmetry-allowed, while the intercalant species tunes its magnitude.

The Ising SOC discussed above, while analyzed in the $k_z = 0$ plane, persists throughout the BZ. At $k_z = 1/2$, the band structure additionally reveals topological nodal-line features. Fig.~\ref{fig:nodal} shows the band structure along the $L$-$H$-$A$ path, the 3D band dispersion, and the (001) surface states for SnNbSe$_2$. The nodal lines, marked by red dots in the band structure and red loops in the 3D dispersion, arise from band crossings protected by the horizontal mirror symmetry $\sigma_h$ of $D_{3h}$. When SOC is included, the relevant symmetry becomes the spinful mirror operator $M_Z=\sigma_h\otimes(i\sigma_Z)$. The Ising SOC term $(\bm{\sigma}\cdot\hat{\mathbf{z}})$ couples crossing bands with opposite $M_z$ eigenvalues, opening a gap; where the two crossing bands share the same $M_z$ eigenvalue, however, this coupling vanishes by symmetry, leaving the crossing intact \cite{engstrom2025}. Along the $L$-$H$-$A$ segment, the crossing bands carry the same $M_z$ eigenvalue, so this portion of the nodal line survives, analogous to the mechanism in PbTaSe$_2$ \cite{bian2016}. The corresponding (001) surface states, calculated for both Sn- and Se-terminations (Figs.~\ref{fig:nodal}(c) and (d)), disperse within the projected bulk gap as drumhead-like bands, a hallmark of bulk nodal lines that is directly accessible by ARPES.

\begin{figure*}[t]
\centering
\includegraphics[width=0.85\linewidth]{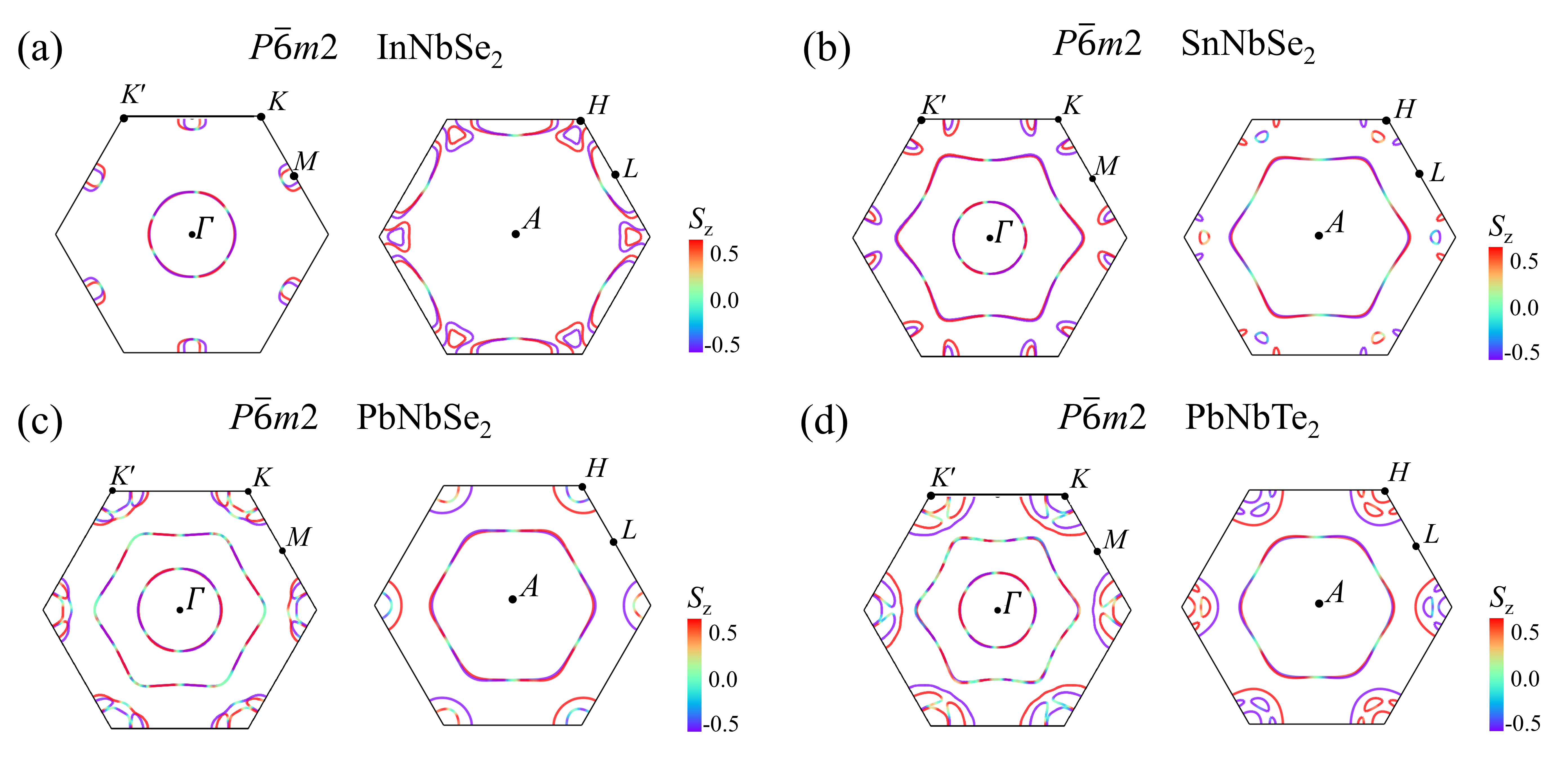}
\caption{Spin-projected Fermi surfaces of the four superconducting candidates in the $P\bar{6}m2$ structure, colored by the out-of-plane spin component $S_z$ (red: positive, blue: negative). For each compound, the left and right columns show the Fermi surfaces at $k_z = 0$ and $k_z = 1/2$ planes, respectively. (a) InNbSe$_2$, (b) SnNbSe$_2$, (c) PbNbSe$_2$ and (d) PbNbTe$_2$.}
\label{fig:fermi}
\end{figure*}

The same nodal-line features and surface states are consistently reproduced across the other three superconducting $P\bar{6}m2$-phase candidates (supplemental Fig.~S5), confirming that topological nodal lines are a robust property of the non-centrosymmetric intercalated structure, independent of the specific intercalant or chalcogen species. The coexistence of nodal-line topology and superconductivity in this structure class is analogous to that in PbTaSe$_2$ \cite{bian2016,zhang2016} and aligns with recent high-throughput predictions of superconducting topological semimetals in intercalated AMX$_2$ compounds \cite{song2024}. These developments position the intercalated $P\bar{6}m2$-phase family at the confluence of band topology and field-robust 3D Ising superconductor.

Having established the electronic structure, SOC, and topological features of the $P\bar{6}m2$ compounds, we now examine their spin texture. The definitive signature of Ising superconductivity is not merely the magnitude of SOC splitting, but the character of the spin polarization, specifically, out-of-plane spin locking at the $K$ and $K'$ valleys with opposite signs. To confirm that this SOC splitting is accompanied by the requisite Ising spin texture, in Fig.~\ref{fig:fermi} we show the spin-projected FS of the four superconducting candidates (InNbSe$_2$, SnNbSe$_2$, PbNbSe$_2$, and PbNbTe$_2$) in the $P\bar{6}m2$ structure, with the out-of-plane spin component $\langle S_z\rangle$ color-coded (red: positive, blue: negative) at $k_z = 0$ and $k_z = 1/2$ planes. For all four compounds, the FS sheets centered around $K$ and $K'$ carry opposite $\langle S_z\rangle$. The persistence of dominant out-of-plane spin polarization at both $k_z = 0$ and $k_z = 1/2$ planes confirms that spin-valley locking is robust throughout the BZ, distinguishing these materials as genuinely 3D Ising superconductors.

\begin{figure*}[t]
\centering
\includegraphics[width=0.85\linewidth]{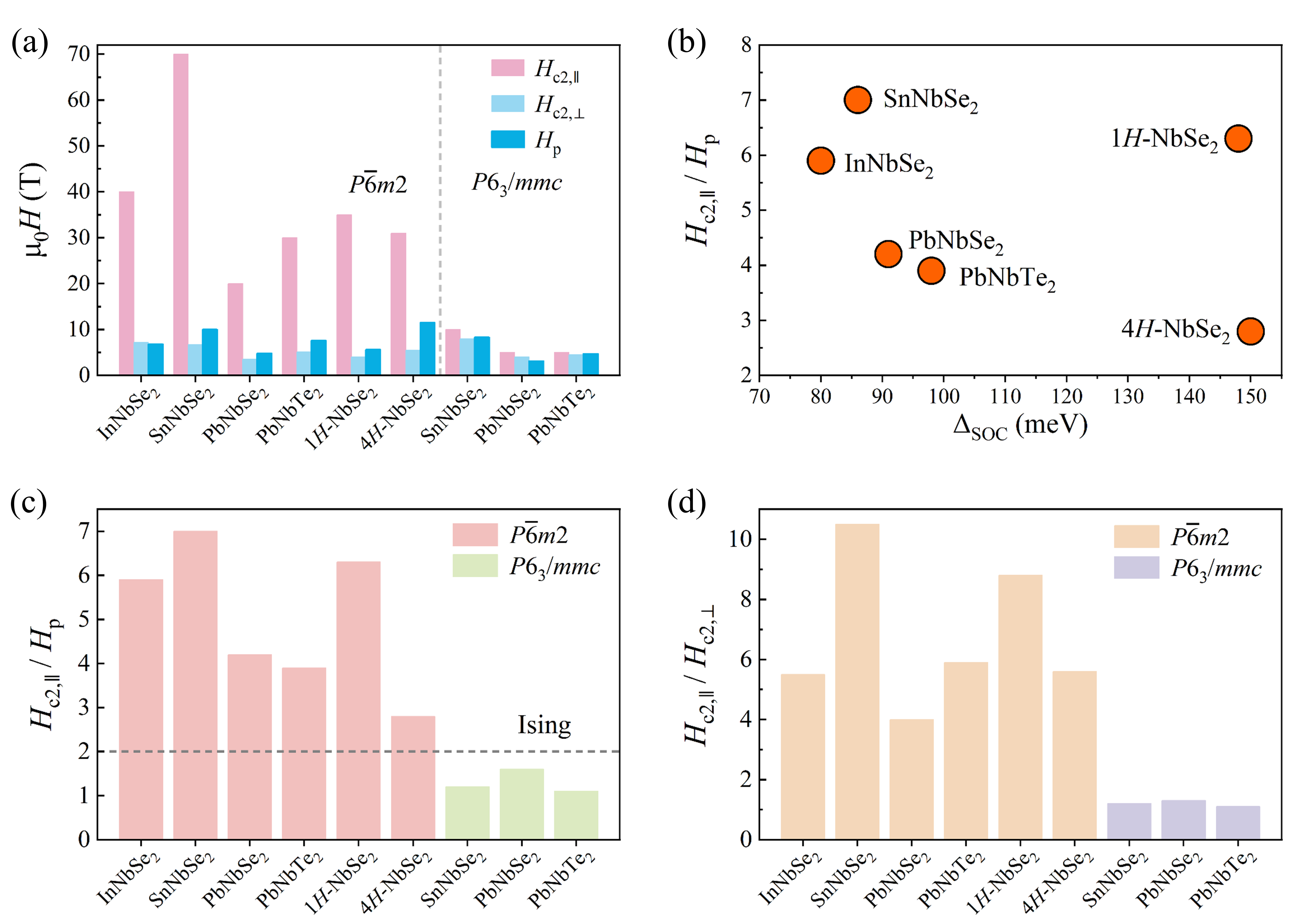}
\caption{(a) Calculated in-plane upper critical field $H_{c2,\parallel}$ (pink), out-of-plane upper critical field $H_{c2,\perp}$ (light blue), and Pauli paramagnetic limit $H_P$ (dark blue) for nine compounds. The vertical dashed line separates the non-centrosymmetric $P\bar{6}m2$ phase (left) from the centrosymmetric $P6_3/mmc$ phase (right). (b) Ratio $H_{c2,\parallel}/H_P$ as a function of the SOC splitting $\Delta_{\mathrm{SOC}}$ for the four compounds in the $P\bar{6}m2$ phase and the two reference non-centrosymmetric polytypes. (c) Ratio $H_{c2,\parallel}/H_P$ for all nine compounds, with the $P\bar{6}m2$ (pink) and $P6_3/mmc$ (green) phases distinguished. The dashed line marks $H_{c2,\parallel}/H_P = 2$. (d) Anisotropy ratio $H_{c2,\parallel}/H_{c2,\perp}$, with the $P\bar{6}m2$ (orange) and $P6_3/mmc$ (purple) phases distinguished.}
\label{fig:fields}
\end{figure*}

While out-of-plane polarization is necessary for Ising protection, the realized degree of protection also depends on the purity of this polarization, quantified by the residual in-plane spin components $S_{x,y}$ that couple to an in-plane field and provide a parasitic depairing channel. Supplemental Fig.~S6 displays the in-plane spin textures for all four candidates. In all cases, $S_{x,y}$ is substantially weaker than $S_z$, confirming their dominant Ising character. However, a quantitative difference emerges: the Pb-intercalated compounds exhibit a quantitatively larger $S_{x,y}$ on the FS than their In- and Sn-intercalated counterparts, despite having larger $\Delta_{\mathrm{SOC}}$. This larger in-plane admixture provides a stronger residual coupling of the Zeeman field to the Cooper pair spins, partially offsetting the Ising protection conferred by its larger bare SOC. This spin-texture-driven depairing is key to understanding the critical-field results that follow.

\begin{figure*}[t]
\centering
\includegraphics[width=0.85\linewidth]{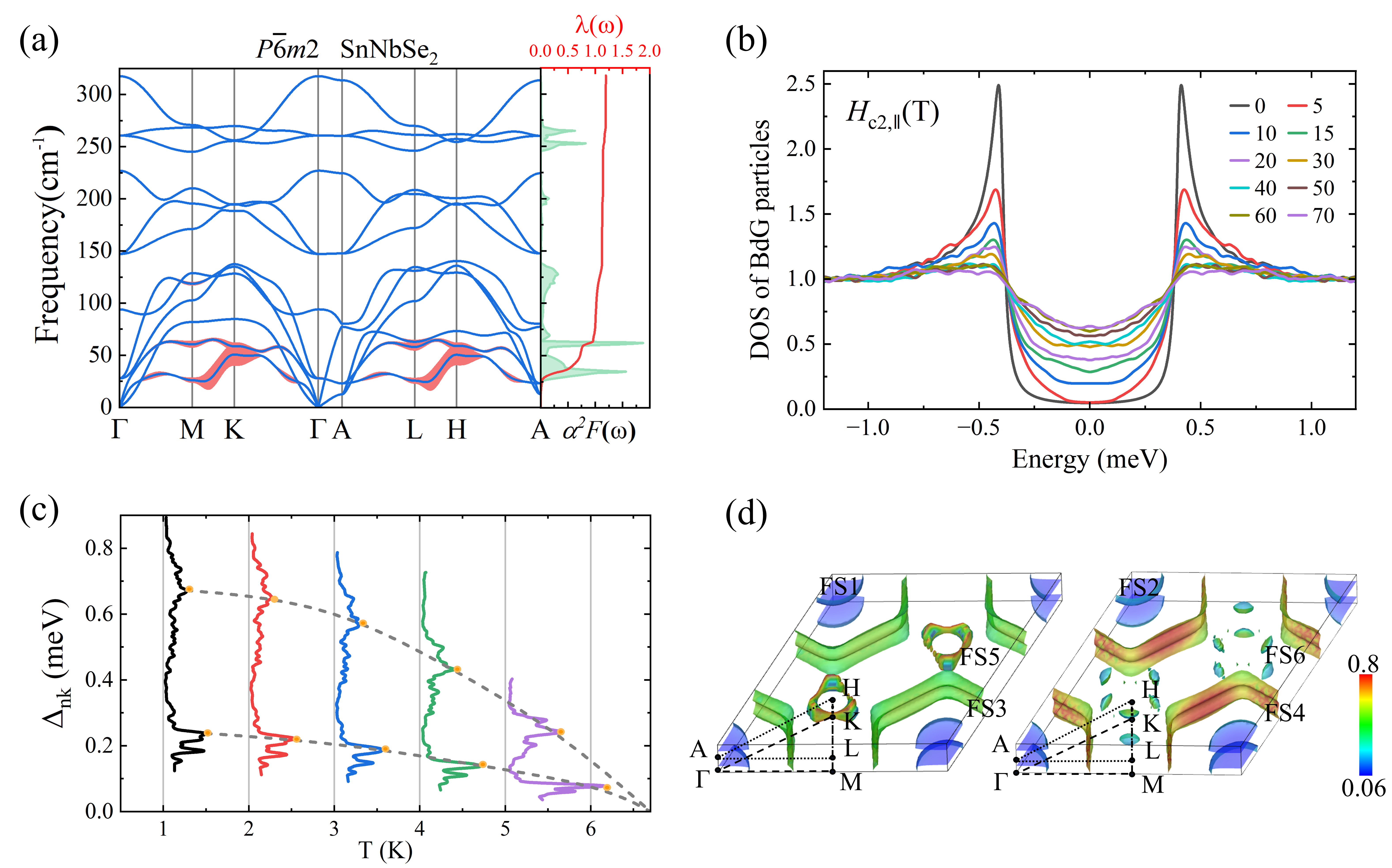}
\caption{(a) Phonon dispersions of SnNbSe$_2$ in the $P\bar{6}m2$ phase weighted by the magnitude of electron-phonon coupling (EPC) strength $\gamma_{\mathbf{q}\nu}$. The Eliashberg spectral function $\alpha^2F(\omega)$ (gray) and the integrated strength of the EPC constant $\lambda(\omega)$ (red) are plotted on the right side. (b) Quasiparticle density of states from the BdG Hamiltonian under different in-plane Zeeman fields. (c) Temperature-dependent superconducting gap $\Delta_{n\mathbf{k}}$. The dashed lines are guides to the eye. (d) Momentum-resolved superconducting gap $\Delta_{\mathbf{k}}$ at $T = 1$ K projected onto the Fermi surface.}
\label{fig:epc}
\end{figure*}

The preceding sections have established that the four candidates in the $P\bar{6}m2$ structure possess the requisite normal-state properties required for Ising superconductivity: broken inversion symmetry, substantial SOC splitting, out-of-plane spin polarization, and finite $T_c$. We now connect these properties to measurable critical fields using a minimal BdG model. For comparison, we also include the three superconducting variants in the $P6_3/mmc$ structure, which have zero Ising SOC splitting, to isolate the role of spin-valley locking in the critical-field enhancement. To estimate the superconducting critical fields, we construct a BdG Hamiltonian in the full Wannier orbital-spin basis following the approach of Ref.~\cite{zheng2025}. A Wannier-based tight-binding Hamiltonian $H(\mathbf{R})$ is constructed by fitting the DFT band structure using Wannier90 \cite{mostofi2014}, retaining the Nb-$4d$, chalcogen-$p$, and intercalant-$p$ orbitals. $H(\mathbf{k})$ is constructed by Fourier interpolation from the real-space hopping matrices $H(\mathbf{R})$:
\begin{equation}
H(\mathbf{k}) = \sum_{\mathbf{R}} H(\mathbf{R})\, e^{2\pi i \mathbf{k}\cdot\mathbf{R}} .
\end{equation}
The BdG Hamiltonian in the Nambu basis is then
\begin{equation}
H_{\mathrm{BdG}}(\mathbf{k}) = \begin{pmatrix} H(\mathbf{k}) + H_Z - \mu I & i\Delta_0\,\sigma_y\otimes I_{\mathrm{orb}} \\ -i\Delta_0\,\sigma_y\otimes I_{\mathrm{orb}} & -H^{*}(-\mathbf{k}) - H_Z^{*} + \mu I \end{pmatrix},
\end{equation}
where the upper-left block is the electron sector with Zeeman coupling $H_Z=(g\mu_B/2)\,\bm{\sigma}\cdot\mathbf{B}\otimes I_{\mathrm{orb}}$ ($g=2$) and chemical potential $\mu$. The pairing block describes $s$-wave spin-singlet pairing with $\Delta_0$ an intra-orbital $s$-wave pairing potential, and $I$ the identity matrix in the orbital basis. The pairing gap is taken as $\Delta_0=1.76\,k_B T_c$ (Table~\ref{tab:table2}). For the EPC constants in this work ($\lambda = 0.58$-$0.98$), strong-coupling corrections to the Bardeen-Cooper-Schrieffer (BCS) gap ratio $2\Delta_0/k_B T_c$ remain modest \cite{carbotte1990}. While broken inversion symmetry can in principle admit mixed pairing channels, the minimal $s$-wave singlet model is sufficient to capture the Ising protection physics that governs the in-plane critical field, and the lower-right hole block is the particle-hole conjugate of the electron sector. The BdG matrix, of dimension $2N\times2N$ where $N$ is the number of Wannier functions, is diagonalized on a uniform $500\times500\times50$ k-point mesh, and the quasiparticle DOS $N(E)$ is accumulated via Lorentzian broadening:
\begin{equation}
N(E) = \frac{1}{N_k}\sum_{\mathbf{k},n} \frac{\eta}{\pi(E - E_{\mathbf{k},n})^2 + \eta^2},
\end{equation}
where $E_{\mathbf{k},n}$ are the BdG eigenvalues, and $\eta$ is the Lorentzian broadening. The upper critical field is identified as the field at which the superconducting coherence peaks in the DOS are suppressed. Fig.~\ref{fig:epc}(b) shows the quasiparticle DOS for SnNbSe$_2$ at selected field values, illustrating the characteristic Ising-protected gap filling: the coherence peaks persist well beyond the Pauli limit and are suppressed only near 70 T ($7\,H_P$). In contrast, under an out-of-plane field, the coherence peaks are suppressed rapidly near 7 T. The DOS evolution under in-plane fields for the other three candidates in the $P\bar{6}m2$ structure is provided in supplemental Fig.~S7, where the same Ising-protected behavior is consistently reproduced. Table~\ref{tab:table2} compiles the resulting critical fields for all seven superconducting systems together with two reference NbSe$_2$ polytypes.

To place our intercalated systems in the broader context of Ising superconductivity, we compare their critical-field properties with the two experimentally known NbSe$_2$ polytypes. The 1$H$ monolayer ($P\bar{6}m2$) is the prototypical 2D Ising superconductor with a K-point valence-band splitting of $\sim$100 meV and $H_{c2,\parallel}/H_P \approx 6.3$ \cite{xi2015}. The 4$H_a$ bulk polytype ($P\bar{6}m2$), recently shown to host 3D Ising superconductivity with $H_{c2,\parallel}/H_P \approx 2.8$ \cite{patra2025,volavka2026}, shares the same space group as our intercalated compounds. Fig.~\ref{fig:fields}(a) compares the in-plane upper critical field $H_{c2,\parallel}$, the out-of-plane upper critical field $H_{c2,\perp}$, and the Pauli paramagnetic limit $H_P$ for all nine compounds as bar charts (Table~\ref{tab:table2}). For the $P\bar{6}m2$ systems, $H_{c2,\parallel}$ clearly exceed $H_P$, with $H_{c2,\parallel}/H_P$ ranging from 3.9 to 7.0, meanwhile, for the $P6_3/mmc$ structure, this ratio remains close to unity ($H_{c2,\parallel}/H_P \approx 1.1$-$1.6$). Fig.~\ref{fig:fields}(b) plots $H_{c2,\parallel}/H_P$ against $\Delta_{\mathrm{SOC}}$ for the four $P\bar{6}m2$ candidates together with monolayer and 4$H_a$-NbSe$_2$. A simple correlation between $\Delta_{\mathrm{SOC}}$ and $H_{c2,\parallel}/H_P$ was not observed. Here the Pb-intercalated compounds exhibit lower enhancement than their In- and Sn-intercalated counterparts despite having a larger SOC splitting. As discussed before, the larger residual in-plane spin component in Pb-intercalated compounds (supplemental Fig.~S6) partially counteracts the Ising protection conferred by its larger $\Delta_{\mathrm{SOC}}$. Fig.~\ref{fig:fields}(c) isolates this contrast with a direct comparison of $H_{c2,\parallel}/H_P$ for all nine systems. A ratio exceeding 2 (dashed line in Fig.~\ref{fig:fields}(c)) is generally taken as the threshold for Ising superconductivity \cite{volavka2026}. This ratio exceeds 2 for every $P\bar{6}m2$ compound (2.8-7.0) and falls below 2 for every $P6_3/mmc$ compound (1.1-1.6). Fig.~\ref{fig:fields}(d) presents the anisotropy ratio $H_{c2,\parallel}/H_{c2,\perp}$, with the $P\bar{6}m2$ systems showing pronounced in-plane enhancement (5.5-10.5) contrasted with the nearly isotropic $P6_3/mmc$ phase (1.1-1.3). Together with the pronounced in-plane enhancement of $H_{c2,\parallel}$ beyond $H_P$, this large anisotropy is a defining characteristic of 3D Ising superconductivity \cite{volavka2026}.

Having characterized the Ising protection and critical-field enhancement, we now examine the EPC that drives superconductivity in these compounds. We take SnNbSe$_2$ as the representative case. Fig.~\ref{fig:epc}(a) shows the phonon dispersion of SnNbSe$_2$ weighted by the EPC strength $\gamma_{\mathbf{q}\nu}$, together with $\alpha^2F(\omega)$ and the cumulative $\lambda(\omega)$. The dominant EPC contribution, visible as the largest $\gamma_{\mathbf{q}\nu}$ weights, is concentrated on the soft phonon branches along the $M$-$K$ and $L$-$H$ paths. The corresponding peak in $\alpha^2F(\omega)$ and the rapid rise of $\lambda(\omega)$ in the same frequency range indicate that modes below $\sim$60 cm$^{-1}$ contribute the dominant share of the EPC strength. The total EPC constant is $\lambda = 0.98$ for SnNbSe$_2$, consistent with previous studies emphasizing the dominant role of EPC in NbSe$_2$-based Ising superconductors \cite{das2023,zheng2019}. The logarithmic average frequency $\omega_{\log} = 78.0$ K and the Allen-Dynes $T_c = 5.4$ K. The EPC parameters for all seven superconducting systems are compared in Table~\ref{tab:table1}.

The previous electronic structure analysis reveals that the compounds in the $P\bar{6}m2$ structure possess multiple FS sheets, predominantly hole pockets at $\Gamma$ and electron pockets around $K$, with distinct SOC strength. In multi-band superconductors, inequivalent FS can develop different superconducting gaps, as exemplified by MgB$_2$ \cite{choi2002}, and recently demonstrated in bulk 4$H_a$-NbSe$_2$ \cite{patra2025,zhou2023}. To determine whether such a multi-gap scenario is realized here, the momentum-resolved gap $\Delta_{n\mathbf{k}}$ was obtained on dense k-point grids at different temperatures. As can be seen in Fig.~\ref{fig:epc}(c), at the lowest temperature ($T = 1$ K), two well-separated peaks are clearly resolved: a larger gap centered at 0.67 meV, corresponding to the pocket along the $M$-$K$ contribution, while a smaller gap centered at 0.24 meV, originating from the $\Gamma$ pocket. As temperature increases, both gaps shift to lower energies and vanish near 6.7 K, yielding an independent estimate of the transition temperature from the anisotropic ME theory, and fall within the range of 5-8 K reported experimentally for Sn-intercalated NbSe$_2$ \cite{munir2021}. Among the other three $P\bar{6}m2$ candidates, PbNbSe$_2$ also exhibits a clear two-peak structure, while InNbSe$_2$ and PbNbTe$_2$ show no clear two-gap features (see supplemental Fig.~S8(d), (e), and (f)).

To identify the origin of this gap anisotropy, Fig.~\ref{fig:epc}(d) shows the calculated momentum-resolved superconducting gap $\Delta_{\mathbf{k}}$ projected onto the FS for SnNbSe$_2$. The gap is finite over all FS sheets, indicating a fully gapped state. This pronounced anisotropy, reproduced across all four $P\bar{6}m2$ candidates (supplemental Fig.~S8(g), (h), and (i)), correlates with the SOC distribution on the FS: the $M$-$K$ pockets, where the SOC splitting is maximal, carry a larger gap, while the $\Gamma$ sheets, with weaker SOC, exhibit a smaller gap. These results demonstrate that anisotropic superconductivity is a generic feature of the intercalated $P\bar{6}m2$ family, with the degree of anisotropy ranging from single-gap to well-resolved two-gap, varying across the series. The anisotropic gap structure also has direct implications for the critical-field analysis: Since the pockets along $M$-$K$ combine a larger gap with stronger out-of-plane SOC, they are expected to dominate the Ising protection, while the $\Gamma$ pocket, with its smaller gap and weaker SOC, likely contributes a subdominant depairing channel at high in-plane fields.

\section{Conclusion}
\label{sec:conclusion}

We have systematically designed 16 intercalated NbSe$_2$- and NbTe$_2$-based compounds, identifying four non-centrosymmetric $P\bar{6}m2$ candidates: InNbSe$_2$, SnNbSe$_2$, PbNbSe$_2$, and PbNbTe$_2$ as viable 3D Ising superconductors. Each exhibits substantial SOC splittings near the $E_F$ of 80-98 meV and $T_c = 2.6$-$5.4$ K. The Ising protection efficiency is governed by out-of-plane spin purity on the FS, not only by the SOC splitting magnitude. Anisotropic Eliashberg calculations reveal strong gap anisotropy, with SnNbSe$_2$ and PbNbSe$_2$ displaying clear two-gap structure linked to the inequivalent FS sheets. BdG calculations predict $H_{c2,\parallel}$ reaching 4-7 times the $H_P$. These findings establish intercalation as a symmetry engineering strategy for realizing 3D Ising superconductors in TMDs and provide both candidate materials and a screening framework to guide future experiments.

\begin{acknowledgments}
This work was supported by the National Key Research and Development Program of China under Contract No.~2022YFA1403200.
\end{acknowledgments}

\end{document}